\documentclass[superscriptaddress,nofootinbib,preprint]{revtex4}
\usepackage{graphicx}
\usepackage[hypertex]{hyperref}

\newcommand{\lsim}{\lesssim}

\newcommand{\hin}{h_{inv}}
\newcommand{\pt}{p_T\!\!\!\!\!\! /\,\,}
\begin{document}

\pagestyle{plain}

\hfill$\vcenter{
\hbox{\bf MADPH-04-1408}
\hbox{\bf FERMILAB-PUB-04-371-T}
                 \hbox{\bf hep-ph/0412269}}$
\vskip 1cm

\title{\mbox{Discovering an Invisibly Decaying Higgs at Hadron Colliders}}

\author{Hooman Davoudiasl}
\email{hooman@physics.wisc.edu}
\affiliation{Department of Physics, University of Wisconsin,
Madison, WI 53706}

\author{Tao Han}
\email{than@physics.wisc.edu}
\affiliation{Department of Physics, University of Wisconsin,
Madison, WI 53706}

\affiliation{Fermi National Accelerator Laboratory, P.O. Box 500, 
Batavia, IL 60510}

\author{Heather E.\ Logan}
\email{logan@physics.wisc.edu}
\affiliation{Department of Physics, University of Wisconsin,
Madison, WI 53706}


\begin{abstract}

A Higgs boson lighter than $2 m_W$ that 
decays mostly into invisible channels ({\it e.g.}, dark matter particles) 
is theoretically well-motivated.  We study the prospects for 
discovery of such an invisible Higgs, $\hin$, at the LHC and the Tevatron 
in three production modes: (1) in association with a $Z$, 
(2) through Weak Boson Fusion (WBF), and (3) accompanied by a jet.  
In the $Z + \hin$ channel, we show that
the LHC can yield a discovery signal above $5\sigma$ with 10 fb$^{-1}$ of 
integrated luminosity for a Higgs mass of 120 GeV.  With 30 fb$^{-1}$
the discovery reach extends up to a Higgs mass of 160 GeV.  
We also study the extraction of the $\hin$ mass from production cross
sections at the LHC, and find that combining WBF and $Z+\hin$ allows
a relatively model-independent determination of the $\hin$ mass with 
an uncertainty of 35--50 GeV (15--20 GeV) with 10 (100) fb$^{-1}$.
At the Tevatron, a 3$\sigma$ observation of a 120~GeV $\hin$ in any 
single channel is not possible with less than 12 fb$^{-1}$ per detector.  
However, we show that combining the signal from WBF with the 
previously-studied 
$Z + \hin$ channel allows a 
3$\sigma$ observation of $\hin$ with 7 fb$^{-1}$ per detector.  
Because of overwhelming irreducible 
backgrounds, $\hin + j$ is not a useful search channel at either the 
Tevatron or the LHC, despite the larger production rate.

\end{abstract}
\maketitle

\section{Introduction}

The Higgs particle is the only missing part of the highly successful Standard 
Model (SM) of particle physics.  The current experimental data 
from direct searches \cite{HiggsDirect} and electroweak precision measurements
\cite{ElectroweakPrecision} point to a Higgs mass in the range 
114~GeV~$< m_h \lsim 250$~GeV.   Thus, if the Higgs exists the Tevatron might 
detect it in the next several years and the LHC is expected to discover it.  

Most analyses assume that the Higgs will predominantly decay into detectable 
SM fields.  However, this may not be a good assumption if there are new weakly 
interacting particles with mass less than half the Higgs mass that couple to 
the Higgs with ${\cal O}(1)$ strength.  In this case, if 
$m_h < 160$~GeV $ \simeq 2 \, m_W$ so that the Higgs partial width
into SM particles is very small, the Higgs will decay predominantly into
the new weakly interacting particles.  In particular, if these new
weakly interacting particles are neutral and stable, the Higgs will decay 
{\it invisibly}.  There are many models in which this situation is 
realized, such as the Minimal Supersymmetric Standard Model 
(MSSM, with Higgs decays to lightest neutralinos), 
models with extra dimensions (with Higgs decays to Kaluza-Klein 
neutrinos \cite{Arkani-Hamed:1998vp}), and Majoron 
models \cite{Joshipura:1992ua}.  An invisible Higgs is also quite 
generic in minimal models of dark matter containing a stable singlet 
scalar \cite{McDonald:1993ex,Burgess:2000yq,Davoudiasl:2004be}.  
From a phenomenological point of view, the existence of dark matter in 
the universe provides compelling evidence for 
stable neutral particles with weak scale masses and couplings.  Given 
unsuppressed couplings and suitable masses, the Higgs would decay nearly 
exclusively into these particles and become invisible in collider 
experiments.  The combined LEP experimental bound on the mass of an 
invisibly-decaying Higgs boson is 114.4~GeV at 95$\%$ confidence 
level \cite{Josa:2001fn}.   

In this paper, we study the discovery potential for the invisible Higgs 
$h_{inv}$ at the LHC and the Tevatron.  We focus on three production 
channels: $Z + \hin$, $\hin + j j $ in Weak Boson Fusion (WBF), and 
$\hin + j$ in gluon fusion.  There have been a number of similar studies in 
the past \cite{Gunion:1993jf,Choudhury:1993hv,Frederiksen:1994me,
Martin:1999qf,Eboli:2000ze,Godbole:2003it,Battaglia:2004js,Belotsky:2004ex}.
We also examine the prospects for determining the mass of the invisible Higgs
from production cross sections at the LHC.  We will show that the $Z + \hin$ 
channel gives a surprisingly good handle on the Higgs mass given 100 fb$^{-1}$
of integrated luminosity.  We will also show how the $Z + \hin$ and WBF 
channels can be combined at the LHC to remove model assumptions from the
Higgs mass extraction.

Discovery of the Higgs in the $Z + \hin$ channel was studied for the LHC in 
Refs.~\cite{Frederiksen:1994me,Godbole:2003it}.
This channel was also analyzed 
for the Tevatron in Ref.~\cite{Martin:1999qf}.  In 
Ref.~\cite{Frederiksen:1994me}, the $Z +$jet background at the LHC was found 
to diminish the significance of the signal considerably, and the 
electroweak backgrounds coming from $W W$ and $ZW$ final states were ignored.  
We will show that, with the kinematic acceptance and the cuts we adopt,
the prospects for the discovery of the invisible Higgs 
in $Z+\hin$ at the LHC are brighter than presented in
Ref.~\cite{Frederiksen:1994me}, even with the $WW$ and $ZW$ backgrounds
included.  Our results are consistent with those of Ref.~\cite{Godbole:2003it}.

WBF production of the invisible Higgs was studied for the LHC in 
Ref.~\cite{Eboli:2000ze}, which showed that WBF can provide significant
signals for invisible Higgs discovery, even at low luminosity.  
Here, we will use their approach to show that WBF contributes significantly
to the observation of $\hin$ at the Tevatron.  Even though a 
3$\sigma$ observation of a 
120~GeV $\hin$ in any single channel at the Tevatron  
is not possible with less than 12 fb$^{-1}$ per detector,
one can enhance the significance of the 
signal by combining data from various channels.  At the Tevatron, an 
important production mode is $Z+\hin$
\cite{Martin:1999qf} and yields a somewhat larger significance 
than the WBF 
channel that we study.  Combining these two channels and data from two 
Tevatron detectors, we show that a 3$\sigma$ observation of $\hin$ with 
$m_h = 120$~GeV can be obtained with 7 fb$^{-1}$ of integrated luminosity 
per detector. 

In the case of $\hin + j$, we study the size of the irreducible background 
generated by $Z (\to \nu {\bar \nu}) + j$.  Although $\hin + j$ is the 
leading triggerable production cross section for $\hin$ at both the LHC
and the Tevatron~\cite{Field:2003yy}, the $Z (\to \nu {\bar \nu}) + j$
background is so large that we conclude that this channel cannot help
to discover $\hin$.

In the next section, we consider the $Z+\hin$ production mode at the 
LHC, present our kinematic cuts, and examine various backgrounds.  
We also discuss the Higgs mass extraction from cross section measurements.
In Sec.~\ref{sec:WBF},
we consider $\hin$ production via WBF at the Tevatron. The contribution 
of the $\hin + j$ mode at both the Tevatron and the LHC is discussed in 
Sec.~\ref{sec:hplusjet}.  Section~\ref{sec:conclusions} 
contains a discussion of our results and other concluding remarks.

\section{Associated \boldmath $Z + \hin$ Production at the LHC}

In this section, we consider the production of $\hin$ in association with a 
$Z$ boson
\begin{equation}
  p \, p \to Z(\to \ell^+\ell^-) + \hin \, \, ; \qquad \ell = e, \mu, 
  \label{pptozh}
\end{equation}
at the LHC.  This process was previously studied for the LHC in 
Refs.~\cite{Frederiksen:1994me,Godbole:2003it}.  
We update and refine the analysis of Ref.~\cite{Frederiksen:1994me} by
taking into account sources of background not included in 
that study and considering a wider acceptance range 
for the leptons.
In our analysis, we assume that the Higgs decays 100\% of the time to invisible
final states, and that the production cross section is the same
as in the SM.  Our results can be easily scaled for other invisible branching
fractions or non-SM production cross sections.
Detection of the $Z + \hin$ signal at the Tevatron has been previously 
studied in Ref.~\cite{Martin:1999qf} and we will later mention their 
results for comparison. We will comment on the effects of departure 
from the assumption of a completely invisible Higgs in 
Sec.~\ref{sec:conclusions}.

\subsubsection{Signal for $\hin$}

As the signal is $\ell^+ \ell^- \pt$, the most significant sources of 
background are 
\begin{equation}
  Z (\to \ell^+ \ell^-) Z (\to \nu {\bar \nu}), \qquad  
  W^+ (\to \ell^+ \nu) W^- (\to \ell^- {\bar\nu}), \qquad
  Z (\to \ell^+ \ell^-) W (\to \ell \nu),
  \label{zzwz}
\end{equation}
(with the lepton from the $W$ decay in $ZW$ missed)
and $Z + {\rm jets}$ final states with fake 
$\pt$~\cite{Frederiksen:1994me,Martin:1999qf}.  We simulate the signal and the 
first three backgrounds for the LHC using 
Madgraph~\cite{Madgraph}.  

We start with the following ``minimal cuts'':
\begin{equation}
	p_T(\ell^{\pm}) > 10 \ {\rm GeV}, \qquad \qquad
	|\eta(\ell^\pm)| < 2.5, \qquad \qquad
	\Delta R(\ell^+\ell^-) > 0.4,
\label{mincuts}
\end{equation}
where $\eta$ denotes pseudo-rapidity and $\Delta R$ is the separation 
between the two particles in the detector, 
$\Delta R \equiv \sqrt{(\Delta \eta)^2 + (\Delta \phi)^2}$; 
$\phi$ is the azimuthal angle.  
The electromagnetic calorimeter at both ATLAS \cite{ATLAS} and 
CMS \cite{CMS} covers the range 
$|\eta| < 3$; however, the electron trigger covers only $|\eta| < 2.5$ (2.6) 
at ATLAS (CMS).  The pseudo-rapidity acceptance for dielectrons could be 
expanded by requiring only one electron within $|\eta| < 2.5$ and the other
within $|\eta| < 3$.  Meanwhile, the muon trigger covers 
$|\eta| < 2.2$ (2.1) at
ATLAS (CMS), with muon identification and momentum measurement 
out to $|\eta| < 2.4$.  We require 
$|\eta(\ell^\pm)| < 2.5$ for both leptons, so that the larger acceptance
for dielectron events compensates the smaller acceptance for dimuon events.

Because we will cut on the invariant 
mass of the dilepton pair to keep only
events in which the dileptons reconstruct to the $Z$ mass,
we imitate the effects of LHC detector resolution by smearing the electron
momenta according to
\begin{equation}
	\Delta E/E = {0.1\over \sqrt{E \ ({\rm GeV})} } \oplus 0.5\%,
\end{equation}
with the two contributions added in quadrature.  
This smearing has a negligible effect on our results. 
We have thus applied the same smearing to the final state with muons.

The $W W$ background can be largely eliminated by requiring 
that the $\ell^+ \ell^-$ invariant mass $m_{\ell^+ \ell^-}$ is close to $m_Z$:
\begin{equation}
	|m_{\ell^+ \ell^-}-m_Z|<10 \ {\rm GeV}.
	\label{mllcut}
\end{equation}  
Also, the $\ell^+$ and $\ell^-$ from two different parent $W$ bosons tend 
to be more back-to-back than the leptons in the signal.  We therefore impose 
an azimuthal angle cut on the lepton pair, 
\begin{equation}
	\Delta\phi_{\ell^+ \ell^-}<2.5\ \ {\rm or}\ \ 143^\circ.
	\label{dphicut}
\end{equation}
This cut also eliminates Drell-Yan backgrounds with fake $\pt$ caused by 
mismeasurement of the lepton energies.

\begin{figure}
\resizebox{1.0\textwidth}{!}{\includegraphics{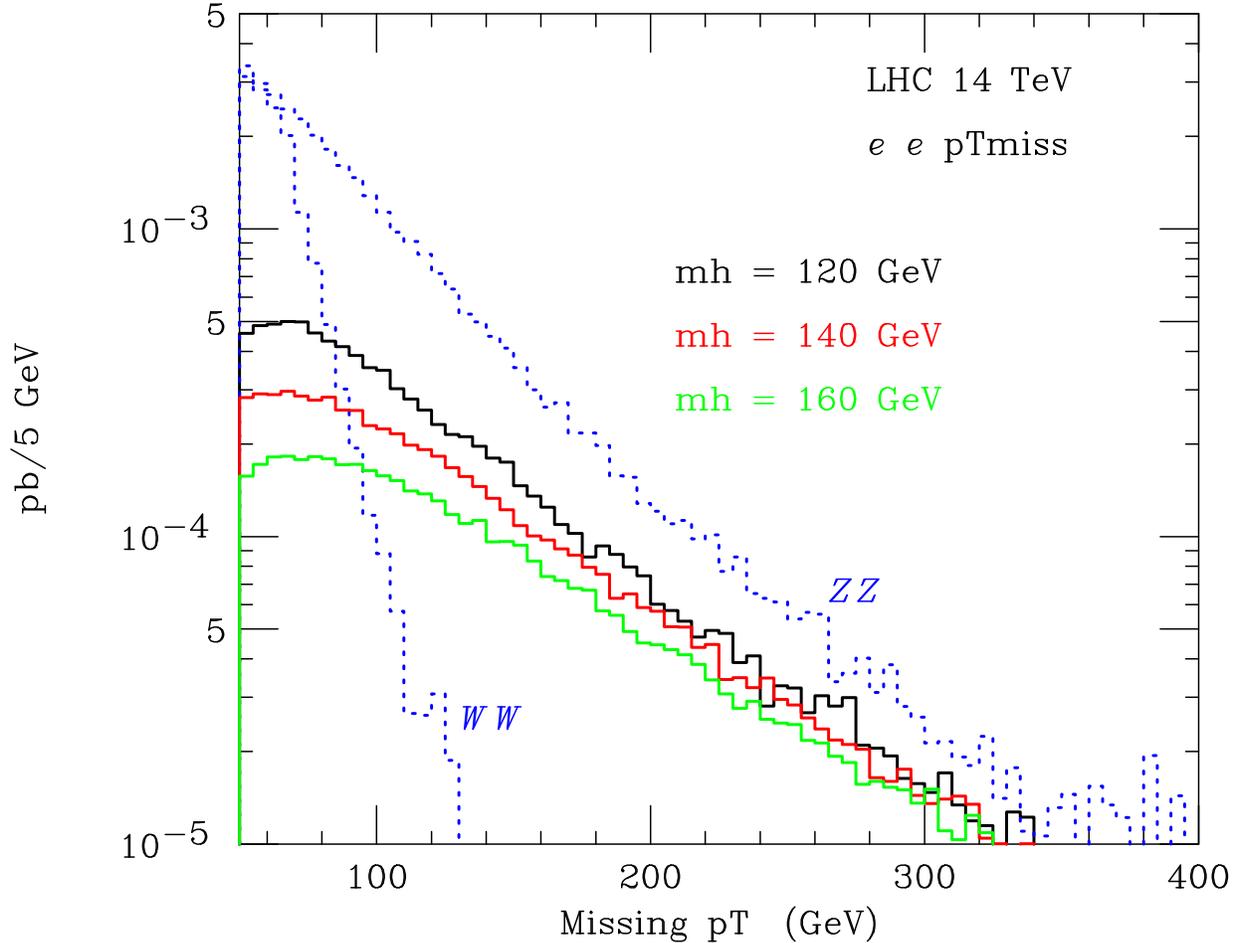}}
\caption{Missing $p_T$ distribution for $Z(\to e^+e^-)+\hin$ signal 
(solid lines, with $m_h = 120$, 140 and 160 GeV top to bottom) 
and backgrounds from $WW$ and $ZZ$ (dotted lines) at the LHC, after
applying the cuts in Eqs.~(\ref{mincuts}), (\ref{mllcut}) and (\ref{dphicut}).
}
\label{fig:ZHptmiss}
\end{figure}

Our third cut is on $\pt$.  The number of $\ell^+ \ell^- \pt$ signal 
events typically falls more slowly with $\pt$ than those of the $ZZ$ 
or $WW$ backgrounds, as shown in Fig.~\ref{fig:ZHptmiss}.  The $\pt$ 
of the $WW$ background is typically 
quite low because the $\pt$ comes from the two neutrinos emitted independently
in the two $W$ decays.  The typical $\pt$ of the $ZZ$ background is somewhat 
larger, but still smaller than that of the signal.  This is because $ZZ$ production
comes from $t$-channel diagrams in which the $Z$ decaying to neutrinos itself
tends to carry less $p_T$ than the $h_{inv}$ produced via
$s$-channel Higgsstrahlung.  As a result, the signal falls 
off with increasing $\pt$ at a slower rate than the $ZZ$ background, 
as reflected by the increase in $S/B$ as $\pt$ gets larger, in Table~\ref{table2}.
The $\pt$ distribution of the signal is also sensitive to the Higgs mass; 
it falls off more slowly with increasing $\pt$, as $m_h$ gets larger.  
Thus a fit to the $\pt$ distribution 
can in principle give some limited sensitivity to the Higgs mass.
We note that the measurement of $\pt$ at hadron colliders suffers from
a lot of systematic effects. However, for the process with $Z\to \ell^+\ell^-$,
the $\pt$ spectrum is largely determined by the well-measured lepton
momenta, which can make the systematic uncertainty minimal.

The final state $Z(\to \ell^+\ell^-) W(\to \ell \nu)$, where the 
lepton from the $W$ decay is missed, can be a potential background.  
However, the probability of
missing the lepton from the $W$ decay is small given the kinematic 
coverage at the LHC.
To reduce this background, we veto events with a third isolated electron
with
\begin{equation}
	p_T > 10 \ {\rm GeV}, \qquad \qquad |\eta| < 3.0.
\end{equation}
For simplicity, we apply the same veto to $W$ decays to muons or taus.
This veto reduces the $Z+W$ background to the level of 5--10 fb, so that
it has little effect on the significance of the signal.  

We also include the background from $Z+{\rm jets}$ with fake $\pt$.
As shown in Ref.~\cite{Frederiksen:1994me}, events of the type 
$Z +$jets can constitute a significant background due to jet energy
mismeasurements resulting in fake $\pt$, or when
one or more jets are emitted outside the fiducial region of the detector
and are therefore missed.  The majority
of those events can be eliminated by applying a jet veto, but those in 
which the jet(s) are soft and/or escape down the beampipe  
can fake $Z+\pt$ events.  A simulation of the latter
requires simulating the detector effects, which is beyond the scope of our 
analysis.  Instead, we adopt the results for this background from 
Ref.~\cite{Frederiksen:1994me}.  We note that these authors 
impose a lepton isolation cone angle of $\Delta R(\ell^+\ell^-) > 0.7$ 
radians, which results in a smaller acceptance than our cuts in 
Eq.~(\ref{mincuts}).  A comparison of our signal and $ZZ$ background
cross sections in Table~\ref{table1} for $\pt > 65$~GeV with those in
the last entry of Table~I in Ref.~\cite{Frederiksen:1994me} 
shows that our larger acceptance
results in a factor of $\sim 1.6$ larger cross sections.

In Ref.~\cite{Frederiksen:1994me}, with the cuts $\pt > 65$~GeV and 
$|m_{\ell^+ \ell^-}-m_Z|<5$~GeV and vetoing jets with $p_T(j) > 45$~GeV 
and $|\eta(j)| < 4.7$, the background cross section coming from $Z +$jets 
is 13.91~fb, combining $e^+e^-$ and $\mu^+\mu^-$ final states.  
Changing the $Z$-mass cut from $|m_{\ell^+ \ell^-}-m_Z|<5$~GeV to 
10~GeV as assumed in our study does not significantly affect the cross
sections of processes with the $\ell^+\ell^-$ coming from a real $Z$ boson.
Rescaling this result by a factor of 1.6 to take into account our larger lepton
acceptance as discussed above, we thus expect a $Z+$jets background 
cross section of $22$ fb for $\pt > 65$~GeV.  
Reference~\cite{Frederiksen:1994me} also found (for a different choice 
of $\eta(j)$ acceptance) that increasing the $\pt$ cut from 65~GeV to 
75~GeV reduces the $Z +$jets background by a factor of about $3.4$.
We thus estimate a $Z +$jets background cross section of about 9~fb for 
$\pt > 75$~GeV with our cuts.  We note that this estimate is quite 
likely a conservative one, since we have only used a factor of 
$\sim 2.5$ reduction of the background in extrapolating from $\pt > 65$~GeV 
to $\pt > 75$~GeV.  In addition, at the LHC, whereas the 
pseudo-rapidity acceptance of the calorimeter in the CMS detector is 
$\pm 4.7$ \cite{CMS}, that of the ATLAS detector is 
$\pm 4.9$ \cite{ATLAS} which helps in the suppression of the fake 
$\pt$ background.

At this point, we note that there are other potentially large sources 
of background that need to be addressed \cite{Martin:1999qf}.  
The background events from $Z^* \to \tau^+ \tau^- \to \ell^+ \ell^- \pt$ 
are efficiently suppressed by our $Z$-mass cut on $m_{\ell^+ \ell^-}$, 
the $\pt$ cut, and the cut on $\Delta \phi_{\ell^+\ell^-}$ that requires 
that the leptons are not back-to-back.  This can be seen from Table 2 in 
Ref.~\cite{Godbole:2003it}, where it is shown that, after cuts similar to those we use, the resulting background from a single $Z$ is basically absent for the $ZH$ production channel.  The same conclusion is reached for the $W + {\rm jet}$ background in the $ZH$ channel, in Table 2 of Ref.~\cite{Godbole:2003it}.  Hence,   
fake events from $W(\to \ell \nu) +$jet, where 
the jet is misidentified as a lepton of the appropriate charge and flavor, 
are also ignored in our analysis.

Our results for the background and signal cross sections are tabulated
in Table~\ref{table1}.  The corresponding signal to background ratio, $S/B$, 
and significance, $S/\sqrt B$,
are tabulated in Table~\ref{table2}.  The numbers given 
in parentheses represent the significance obtained including our estimated 
$Z +$jets background discussed above.  To be cautious, we only consider this 
background for the cases with $\pt > 65$~GeV and $\pt > 75$~GeV, which
were studied in Ref.~\cite{Frederiksen:1994me}, and refrain 
from extrapolating to other $\pt$ cut values.  We see from 
Table~\ref{table2} that a $>5\sigma$ discovery can be obtained for 
$m_h = 120$~GeV with 10 fb$^{-1}$ of integrated luminosity, even with 
our conservative estimate for the $Z+$jets background for $\pt > 75$~GeV.  

\begin{table}[htb]
\begin{tabular}{l|cccc|ccc}
\hline \hline
          &         &         &         &            &
	\multicolumn{3}{c}{S($Z+\hin$)} \\
$\pt$ cut & B($ZZ$) & B($WW$) & B($ZW$) & B($Z+j)^*$ & 
	$m_h = 120$~~~ & 140~~~  & 160 GeV \\
\hline
65 GeV & 48.0 fb & 10.6 fb & 10.2 fb & 22 fb &
	14.8 fb & 10.8 fb & 7.9 fb \\
75 GeV & 38.5 fb & 4.3 fb & 7.4 fb & 9 fb &
	12.8 fb & 9.4 fb & 7.0 fb \\
85 GeV & 30.9 fb & 1.8 fb & 5.5 fb & &
	11.1 fb & 8.3 fb & 6.3 fb \\
100 GeV & 22.1 fb & 0.6 fb & 3.6 fb & &
	8.7 fb & 6.8 fb & 5.3 fb \\
\hline \hline
\end{tabular}
\caption{
Background and signal cross sections for 
associated $Z(\to \ell^+\ell^-)+\hin$ production at the 
LHC, combining the $ee$ and $\mu\mu$ channels.
$^*$Estimated from Ref.~\cite{Frederiksen:1994me}
(see text for details).
}
\label{table1}
\end{table}

\begin{table}[htb]
\begin{tabular}{l|ccc|c|c}
\hline \hline
          & \multicolumn{3}{c|}{$m_h = 120$ GeV} 
		& $m_h = 140$ GeV & $m_h = 160$ GeV \\
$\pt$ cut & S/B & S/$\sqrt{\rm B}$ (10 fb$^{-1}$) 
	& S/$\sqrt{\rm B}$ (30 fb$^{-1}$)
		& S/$\sqrt{\rm B}$ (30 fb$^{-1}$) 
			& S/$\sqrt{\rm B}$ (30 fb$^{-1}$) \\
\hline
65 GeV & 0.22 (0.16) & 5.6 (4.9) & 9.8 (8.5) & 7.1 (6.2) & 5.2 (4.5) \\
75 GeV & 0.25 (0.22) & 5.7 (5.3) & 9.9 (9.1) & 7.3 (6.7) & 5.4 (5.0) \\
85 GeV & 0.29        & 5.7       & 9.8       & 7.4       & 5.6       \\
100 GeV & 0.33       & 5.4       & 9.3       & 7.3       & 5.7       \\
\hline \hline
\end{tabular}
\caption{
Signal significance for
associated $Z(\to \ell^+\ell^-)+\hin$ production at the LHC,
combining the $ee$ and 
$\mu\mu$ channels.  The numbers in the parentheses include the 
estimated $Z + $jets background discussed in the text.  
}
\label{table2}
\end{table}

Reference~\cite{Frederiksen:1994me} finds a $14 \sigma$ signal for 
$\hin$ at the LHC with 100~fb$^{-1}$ and $\pt > 65$~GeV, with the rest
of their cuts as mentioned above.  Rescaling this result for 10~fb$^{-1}$
yields a $4.4 \sigma$ signal, somewhat more pessimistic than our
result for the signal significance for $\pt > 65$~GeV in Table~\ref{table2}.
Our larger significance for $\pt > 65$~GeV is due solely to our larger
lepton acceptance: we accept roughly 1.6 times as many events as the 
study in Ref.~\cite{Frederiksen:1994me}.  Rescaling their results by 
$\sqrt{1.6}$ to account for this larger acceptance, their significance
becomes $5.6 \sigma$.  Our significance for $\pt > 65$ GeV
in Table~\ref{table2} is lower 
than 5.6$\sigma$ because we have included the $WW$ and $ZW$ backgrounds, 
which were neglected
in Ref.~\cite{Frederiksen:1994me}.  We also found that the signal 
significance can be improved somewhat by increasing the $\pt$ cut above
65~GeV; the optimum cut appears to be roughly 75--85~GeV.

Our results are in qualitative agreement with the study of Godbole
et al.\ in Ref.~\cite{Godbole:2003it}, which included hadronization of the
$Z+\hin$ signal and backgrounds using Pythia/Herwig.
For the same cut on $\pt$ of 100 GeV, Ref.~\cite{Godbole:2003it} found a
signal cross section smaller by about 30\% than our result,
and a total background cross section (dominated by $ZZ$ production)
smaller by about 20\%.
We expect that this reduction in both the signal and background cross
sections is due to events being rejected by the jet veto imposed in
Ref.~\cite{Godbole:2003it} after including QCD initial-state radiation.
The 30\% reduction in signal cross section can be compensated
\cite{Martin:1999qf} by the known NLO QCD K-factor for $Z+h$ at LHC of
about 1.3 \cite{zh-nnlo}, yielding a signal cross section consistent
with our result.
Similarly, the reduction in the dominant $ZZ$ background can be
compensated by the known NLO QCD K-factor for $ZZ$ at LHC of
about 1.2 \cite{zz-nlo}, yielding a background cross section
consistent with our result.
We have not explicitly included any K-factors in our signal or background
cross section calculations, with the expectation of some reduction due to
the jet vetoing requirement.

For comparison, a $3 \sigma$ observation of $\hin$ at the same mass at 
the Tevatron will require 26 fb$^{-1}$, and with 30 fb$^{-1}$ it is 
possible to observe a 125~GeV $\hin$ at the $3 \sigma$ level 
\cite{Martin:1999qf}.  However, at the LHC, with 30 fb$^{-1}$ and given 
our conservative estimate of $Z + $jets background,
a $5\sigma$ discovery or better is possible up to $m_h = 160$~GeV, as 
shown in Table~\ref{table2}.  For higher Higgs masses, the $h \to WW$ 
decay goes on-shell, increasing the Higgs width
significantly.  The decay to $WW$ will then most likely compete with the 
invisible decay mode, resulting in a partly-visible Higgs.

The $Z+\hin$ channel can thus be used at the LHC for $m_h \lesssim 160$ GeV
to supplement the WBF channel \cite{Eboli:2000ze}, which has higher 
significance.  
However, we would like to emphasize that the $\pt$ measurements in the
process $\ell^+\ell^- \pt$ that we studied here 
are largely determined by $p_T(\ell\ell)$,
and the distribution will suffer much less from systematic uncertainties
compared to the WBF where $\pt$ is determined mainly from the forward jets.

\subsubsection{Higgs boson mass}

The $Z+\hin$ channel may also provide an interesting handle on
the Higgs boson mass, as follows.
The mass of an invisibly-decaying Higgs boson obviously cannot be 
reconstructed from the Higgs decay products.  Unless the Higgs is also
observed in a visible channel, our only chance of determining the Higgs 
mass comes from the $m_h$ dependence of the production process.
Extracting $m_h$ from the production cross section requires the assumption
that the production couplings are the same as in the SM.  (Non-observation
of the Higgs in any visible final state implies that the invisible branching
fraction is close to 100\%.)

The Higgs mass extraction from measurements of the production cross
sections in $Z+\hin$ and WBF are shown in Tables~\ref{Tab:MH:Z+h} and
\ref{Tab:MH:WBF}, respectively.  There are two sources of uncertainty in the 
signal: statistical and from background normalization.  
The statistical uncertainty is 
$\Delta \sigma_S/\sigma_S = \sqrt{\rm S+B}/{\rm S}$.
We estimate the total background normalization 
uncertainty for 
$Z + \hin$ to be the same size as that of the dominant process involving 
$Z \to \nu \nu$: $\Delta {\rm B}/{\rm B} = \Delta {\rm B}(ZZ)/{\rm B}(ZZ)$.  We assume that this background can be measured via
the corresponding channels in which $Z \to \ell^+ \ell^-$
and take the uncertainty to be the statistical uncertainty 
on the $Z \to \ell^+ \ell^-$ rate:  $\Delta {\rm B}(ZZ)/{\rm B}(ZZ) \simeq 
7.1\% \; (2.2\%)$, 
for an integrated luminosity of 10 (100) fb$^{-1}$.  
In Tables~\ref{Tab:MH:Z+h} and \ref{Tab:MH:WBF} we quote the resulting
uncertainty on the signal cross section, given by  
$\Delta \sigma_S/\sigma_S = ({\rm B}/{\rm S})\times \Delta {\rm B}/{\rm B}$.  The total uncertainty $[\Delta \sigma_S/\sigma_S]_{tot}$, presented in 
Tables~\ref{Tab:MH:Z+h} and \ref{Tab:MH:WBF}, is then the sum, in quadrature, of the statistical and background uncertainties, as well as other uncertainties that may exist.  We then have $\Delta m_h$ = $(1/\rho)[\Delta \sigma_S/\sigma_S]_{tot}$; $\rho$ is defined in Tables~\ref{Tab:MH:Z+h} and
\ref{Tab:MH:WBF}.

\begin{table}
\begin{tabular}{lccc}
\hline \hline
$m_h$ (GeV)
 &   120
 &   140
 &   160 \\
\hline
$\rho = (d\sigma_S/dm_h)/\sigma_S$ (1/GeV)
 &  $-0.013$
 &  $-0.015$
 &  $-0.017$ \\
Statistical uncert.
 &   21\% (6.6\%)
 &   28\% (8.8\%)
 &   37\% (12\%) \\
Background normalization uncert.
 &   33\% (10\%)
 &   45\% (14\%)
 &   60\% (19\%) \\
Total uncert.
 &   40\% (16\%)
 &   53\% (19\%)
 &   71\% (24\%) \\
\hline
$\Delta m_h$ (GeV)
 &   30 (12)
 &   35 (12)
 &   41 (14) \\
\hline \hline
\end{tabular}
\caption{Higgs mass determination from $Z + \hin$ with 10 (100) fb$^{-1}$, 
assuming Standard Model production cross section and 100\% invisible decays.  
The signal and background cross sections were taken from Table~\ref{table1} 
for $\pt > 75$ GeV.  The total uncertainty includes a
theoretical uncertainty on the signal cross section from QCD and PDF 
uncertainties of 7\% \cite{ZHQCD} and an estimated
lepton reconstruction efficiency uncertainty of 4\% (2\% per lepton)
and luminosity normalization uncertainty of 5\% \cite{Duhrssen}.
}
\label{Tab:MH:Z+h}
\end{table}

\begin{table}
\begin{tabular}{lccccccc}
\hline\hline
$m_h$ (GeV)
 &   120
 &   130
 &   150
 &   200 \\
\hline
$\rho = (d\sigma_S/dm_h)/\sigma_S$ (1/GeV)
 &  $-0.0026$
 &  $-0.0026$
 &  $-0.0028$
 &  $-0.0029$ \\
Statistical uncert.
 &   5.3\% (1.7\%)
 &   5.4\% (1.7\%)
 &   5.7\% (1.8\%)
 &   6.4\% (2.0\%) \\
Background normalization uncert.
 &   5.2\% (2.1\%)
 &   5.3\% (2.1\%)
 &   5.6\% (2.2\%)
 &   6.5\% (2.6\%) \\
Total uncert.
 &   11\% (8.6\%)
 &   11\% (8.6\%)
 &   11\% (8.6\%)
 &   12\% (8.8\%) \\
\hline
$\Delta m_h$ (GeV)
 &   42 (32)
 &   42 (33)
 &   41 (31)
 &   42 (30) \\
\hline \hline
\end{tabular}
\caption{Higgs mass determination from ${\rm WBF} \to \hin$ 
with 10 (100) fb$^{-1}$, assuming Standard Model
production cross section and 100\% invisible decays.  The background 
and signal cross sections were taken from Tables II and III, respectively, of 
Ref.~\cite{Eboli:2000ze}, and include a central jet veto.  
The total uncertainty includes a 
theoretical uncertainty from QCD and PDF uncertainties of 4\% \cite{WBFQCD},
and an estimated uncertainty on the efficiency of the WBF jet tag and 
central jet veto of 5\% and luminosity normalization uncertainty of 5\% 
\cite{Duhrssen}.
}
\label{Tab:MH:WBF}
\end{table}

The cross section for $Z+\hin$ production falls quickly with increasing $m_h$ 
due to the $s$-channel propagator suppression.  This is in contrast to the
WBF production, which provides a $>5\sigma$ signal up to 
$m_h \simeq 480$ GeV with 10 fb$^{-1}$ if the 
Higgs decays completely invisibly \cite{Eboli:2000ze}.  
Thus, while the statistics are much better on the WBF measurement than on
$Z + \hin$, the systematic uncertainties hurt WBF more
because $(d\sigma_S/dm_h)/\sigma_S$ is much smaller for WBF than for
$Z + \hin$.  The $Z+\hin$ cross section
is therefore more sensitive to the Higgs mass than the WBF cross section.

More importantly, however, taking the ratio of the $Z+\hin$ and
WBF cross sections allows for a more model-independent determination
of the Higgs mass.  This is due to the fact that the production 
couplings in $Z+\hin$ ($hZZ$) and in WBF (contributions from $hWW$ and 
$hZZ$) are related by custodial SU(2) symmetry 
in any model containing only Higgs doublets and/or singlets.  The production
couplings thus drop out of the ratio of rates in this wide class 
of models (which includes the MSSM, multi-Higgs-doublet models, and
models of singlet scalar dark matter), 
leaving dependence only on the Higgs mass.  The resulting Higgs mass
extraction is illustrated in Table~\ref{Tab:MH:Ratio}.
Assuming SM event rates for the statistical uncertainties, 
we find that the Higgs mass can be extracted with an uncertainty of
35--50 GeV (15--20 GeV) with 10 (100) fb$^{-1}$ of integrated luminosity.
The ratio method also allows a test of the SM cross section assumption by
checking the consistency of the $m_h$ determinations from the 
$Z+\hin$ and WBF cross sections alone with the $m_h$ value
extracted from the ratio method.
Furthermore, observation of the invisibly-decaying Higgs in WBF but not in 
$Z+\hin$ allows one to set a lower limit on $m_h$ in this class of models.

\begin{table}
\begin{tabular}{lccc}
\hline \hline
$m_h$ (GeV)
 &   120
 &   140
 &   160 \\
\hline
$r = \sigma_S(Zh)/\sigma_S({\rm WBF})$
 &   0.132
 &   0.102
 &   0.0807 \\
$(dr/dm_h)/r$ (1/GeV)
 &  $-0.011$
 &  $-0.013$
 &  $-0.013$ \\
Total uncert., $\Delta r/r$
 &   41\% (16\%)
 &   54\% (20\%)
 &   72\% (25\%) \\
\hline
$\Delta m_h$ (GeV)
 &   36 (14)
 &   43 (16)
 &   53 (18) \\
\hline \hline
\end{tabular}
\caption{Higgs mass determination from the ratio method discussed in the
text, with 10 (100) fb$^{-1}$.  
The event rates for WBF were interpolated linearly for Higgs 
masses of 140 and 160 GeV, which were not given explicitly in 
Ref.~\cite{Eboli:2000ze}.  
Statistical uncertainties were obtained assuming SM signal rates.
The total uncertainty includes
theoretical uncertainties from QCD and PDF uncertainties of 7\% for 
$Z + \hin$ \cite{ZHQCD} and 4\% for WBF \cite{WBFQCD},
and estimated uncertainties on the lepton reconstruction efficiency in 
$Z+\hin$ of 4\% (2\% per lepton) and on the efficiency of the WBF jet 
tag and central jet veto of 5\% \cite{Duhrssen}.  
The luminosity normalization uncertainty cancels out in the
ratio of cross sections and is therefore not included.
}
\label{Tab:MH:Ratio}
\end{table}

\section{Production of \boldmath $\hin$ via WBF at the Tevatron}
\label{sec:WBF}

WBF provides a significant Higgs production mechanism at the LHC and is a 
promising channel for studying Higgs couplings to weak bosons 
\cite{Rainwater:1997dg,Rainwater:1998kj,Plehn:2001nj}.  
Reference~\cite{Eboli:2000ze} studied $\hin$ production in WBF at the LHC
and concluded that with only 10 fb$^{-1}$
of integrated luminosity, $\hin$ can be detected at the $\geq 5 \sigma$
level up to $m_h \simeq 480$~GeV.  They also showed that the invisible 
branching fraction of a 120~GeV Higgs can be constrained at the 95\% 
confidence level to be 
less than $13\%$ if no signal is seen in the WBF$\to \hin$ channel, 
again with 10 fb$^{-1}$.  

The kinematic requirements for suppressing the backgrounds rely on the 
large energy 
and rapidity of the forward tagging jets characteristic of WBF at the LHC, 
together with the large rapidity coverage
of the LHC detectors.  However, given the more limited kinematic range
and rapidity coverage at the Tevatron, it is not immediately clear whether 
implementing the same search strategy will yield a 
useful signal for $\hin$.  In the following, we will show that the WBF 
production mode will indeed have a significant impact on the prospects 
for the observation of $\hin$ at the Tevatron, before data from the LHC 
becomes available.  

The signal here is $\pt + 2 j$.  
A large background comes from 
$Z (\to \nu {\bar \nu}) + 2 j$ with the jets produced via QCD.
A smaller, but less reducible, background comes from 
$Z (\to \nu {\bar \nu}) + 2 j$ in which the $Z$ is produced by WBF
and the jets have kinematics similar to that of the signal.
In addition, there are backgrounds from $W(\to \ell \nu) + 2j$, in which
the lepton from the $W$ decay is missed, and
QCD backgrounds with fake $\pt$ from missed jets in 
multi-jet events and jet energy mismeasurements in di-jet events.  

We generate the signal, $\hin + 2 j$, the QCD and electroweak backgrounds
with $Z(\to \nu \bar \nu) + 2 j$, and the QCD background with 
$W(\to \ell \nu) + 2 j$ for the Tevatron using Madgraph \cite{Madgraph}.  
We start with the following ``minimal cuts'':
\begin{equation}
	p_T(j) > 10 \ {\rm GeV}, \qquad \qquad
	|\eta(j)| < 3.0, \qquad \qquad 
	\Delta R(jj) > 0.4, \qquad \qquad 
	\pt > 90 \ {\rm GeV}.
\label{eq:WBFbasiccuts}
\end{equation}
The $\pt > 90$~GeV requirement provides a trigger.  We take the calorimeter
pseudo-rapidity coverage from, {\it e.g.}, Ref.~\cite{DZeroeta}.

In WBF events, the two jets come from the initial
partons, which are at high energy and are not deflected very much by the
interaction.  
To separate the signal from the backgrounds, we thus impose ``WBF cuts'': we 
require that the two jets reconstruct to a large invariant mass,
\begin{equation}
	m_{jj} > 320, \ 340, \ 360, \ 400 \ {\rm GeV},
	\label{eq:mjjcuts}
\end{equation}
and are separated by a large rapidity gap,
\begin{equation}
	\Delta \eta_{jj} > 2.8.
\label{eq:WBFdetacut}
\end{equation}
These two cuts eliminate most of the QCD $Z+2j$ and $W+2j$ backgrounds, 
in which the jets
tend to be softer and have a smaller rapidity gap, while preserving a 
significant fraction of the WBF signal. 

To reduce the $W+2j$ background further, we apply a lepton veto.  We
veto events that contain an isolated electron with \cite{CDFelectronveto}
\begin{equation}
	p_T(\ell) > 8 \ {\rm GeV}, \qquad \qquad |\eta(\ell)| < 3.0.
\end{equation}
For simplicity, we apply the same veto to $W$ decays to muons or taus.
Loosening the veto requirements to 
$p_T(\ell) > 10$ GeV, $|\eta(\ell)| < 2.0$ increases
the $W+2j$ background by about a factor of two.

Background can also come from QCD multi-jet events with fake $\pt$ 
due to mismeasurement of jets and jet activity escaping down the beampipe.
We follow the techniques of a CDF study of $\pt + 2 j$ \cite{Acosta:2004zb} 
to deal with this
background.  First we require that the $\pt$ not be aligned with either
of the jets:
\begin{equation}
        \Delta \phi (j,\pt) > 30^{\circ}.
\label{eq:dphijpt}
\end{equation}
This eliminates backgrounds containing fake $\pt$ due to jet energy 
mismeasurement in two-jet events.

The remaining QCD $jj\pt$ background with fake $\pt$ was simulated in 
Ref.~\cite{Acosta:2004zb} for various minimum $\pt$ cuts.  
The kinematic cuts on the jets used in Ref.~\cite{Acosta:2004zb} were 
different than ours: Ref.~\cite{Acosta:2004zb} required
$E_T(j_1) > 40$ GeV, $E_T(j_2) > 25$ GeV, and $|\eta(j_1)|, |\eta(j_2)| < 1$.
They allowed a third jet with $E_T > 15$ GeV and $|\eta|<2.5$, and vetoed
events with any additional jets with $E_T>15$ GeV and $|\eta|<3.6$.
In addition to requiring that the $\pt$ not be aligned with either
of the jets, Eq.~(\ref{eq:dphijpt}), they required that the two 
central jets not be back-to-back, $\Delta \phi(j_1,j_2) < 165^{\circ}$,
and that the $\pt$ not be antiparallel to the leading jet,
$\Delta \phi(j_1,\pt)<165^{\circ}$, in order to eliminate backgrounds
with fake $\pt$ from jet energy mismeasurements.
For $\pt > 90$ GeV, Ref.~\cite{Acosta:2004zb} found a QCD $jj\pt$ 
background with fake $\pt$ of about 5~fb.

The study in Ref.~\cite{Acosta:2004zb} considers jets in the central
region, $|\eta|<1$, in contrast to our cuts in Eqs.~(\ref{eq:WBFbasiccuts})
and (\ref{eq:WBFdetacut}).  However,
since our cut on the dijet invariant mass, $m_{jj}>320$ GeV or higher,
requires much more visible energy in the jets
than the CDF study does, we expect that the QCD background with
fake $\pt$ found in Ref.~\cite{Acosta:2004zb} represents a 
{\it conservative upper limit} for our cuts.  A 
quantitative estimate of the dijet background 
with fake $\pt$ would require simulating the detector effects, 
which is beyond the scope of our analysis.

In Table~\ref{tab:WBF1} we show results
for signal and background cross sections for the $m_{jj}$ cuts given
in Eq.~(\ref{eq:mjjcuts}).
In Table~\ref{tab:WBF2} we show the resulting signal-to-background ratio and
significance for 10 fb$^{-1}$.

\begin{table}
\begin{tabular}{ccccc}
\hline \hline
$m_{jj}$ cut & S($\hin+2j$) & B($Z+2j$,QCD) & B($Z+2j$,EW) & B($W+2j$,QCD) \\
\hline
320 GeV & 4.1 fb & 55 fb & 1.7 fb & 7 fb \\
340 GeV & 3.6 fb & 43 fb & 1.6 fb & 5 fb \\
360 GeV & 3.2 fb & 34 fb & 1.4 fb & 5 fb \\
400 GeV & 2.4 fb & 21 fb & 1.2 fb & 2 fb \\
\hline \hline
\end{tabular}
\caption{Signal and background cross sections for $\hin+2j$ at Tevatron Run 2,
for $m_h = 120$ GeV.
The statistical uncertainty on B($Z+2j$,QCD) after cuts is roughly 
10\% due to our limited Monte Carlo sample.
There is an additional background from QCD with fake $\pt$ which is
taken from Ref.~\cite{Acosta:2004zb} to be 5 fb; this
represents a conservative overestimate of the fake $\pt$ background.}
\label{tab:WBF1}
\end{table}

\begin{table}
\begin{tabular}{cccc}
\hline \hline
$m_{jj}$ cut & S (10 fb$^{-1}$) & S/B & S/$\sqrt{\rm B}$ (10 fb$^{-1}$) \\
\hline
320 GeV & 41 evts & 0.060 & 1.6 \\
340 GeV & 36 evts & 0.066 & 1.5 \\
360 GeV & 32 evts & 0.070 & 1.5 \\
400 GeV & 24 evts & 0.082 & 1.4 \\
\hline \hline
\end{tabular}
\caption{Number of signal events, signal-to-background ratio, and
significance for $\hin+2j$ at Tevatron Run 2, for $m_h = 120$ GeV.
We include the background from QCD with fake $\pt$ of 5 fb \cite{Acosta:2004zb}
in S/B and S/$\sqrt{\rm B}$.
}
\label{tab:WBF2}
\end{table}

From the numbers given in Table~\ref{tab:WBF2} 
it is clear that even with 10$^{-1}$~fb of integrated 
luminosity, the significance is below 2$\sigma$.  
We find a signal significance of about 1.6$\sigma$ with 10 fb$^{-1}$ 
of luminosity at one Tevatron detector.  This significance is not much less 
than that found in Ref.~\cite{Martin:1999qf}
for $Z+\hin$ at the Tevatron, namely $1.9 \sigma$ with 10~fb$^{-1}$ for
$m_h = 120$~GeV.  Neither of these channels alone is sufficient to provide 
an observation
of $\hin$: combining data from both Tevatron detectors, a $3 \sigma$ 
observation
would require at least 12~fb$^{-1}$ in the $Z+\hin$ channel,
or 18~fb$^{-1}$ in the WBF channel.
However, by combining these two channels, we find that a 3$\sigma$ 
observation of $\hin$ is possible with 7 fb$^{-1}$ per detector,
if the background can be determined to better than 10\%.
Thus, WBF provides an important second channel that brings an observation 
of $\hin$ into the realm of possibility at the Tevatron before the results 
of the LHC become available.  Here, we note that there may be other production channels, such as $g g \to h_{inv} jj$, that could contribute to the signal, even after the WBF cuts we have outlined.  However, this could only enhance $h_{inv}$ production, making our results for the WBF channel a lower bound on the number of signal events. 

The $Z(\to \nu\bar\nu) + 2j$ QCD background could be further reduced
by taking advantage of its different color structure 
compared to the signal process.  An important feature of WBF is the
absence of color exchange between the two forward tagging jets, which
results in less hadronic activity in the rapidity region between these 
jets \cite{centraljetveto}.  
Thus, vetoing additional soft jets in the central
region could significantly reduce the QCD background while preserving 
most of the WBF signal.  In Refs.~\cite{Eboli:2000ze} and \cite{DaveThesis}, it is claimed that such a veto improves the 
signal-to-background ratio by a factor of three at the LHC.
If a similar background reduction could be achieved at the Tevatron,
the prospects for $\hin$ observation in the WBF channel would improve 
considerably: a $3\sigma$ observation in the WBF channel alone would then
be possible with 6 fb$^{-1}$ per detector, with a signal-to-background 
ratio close to 1/5.

It is also important to consider the background normalization.
In particular, to better understand the $\pt$ distribution of the 
$Z(\to \nu\bar\nu) + 2j$ backgrounds, one may be able to make use of 
the channel $Z(\to \ell^+\ell^-)+2j$ with the $p_T$ of the $Z$ boson 
reconstructed from the momenta of the two jets (to duplicate the systematic
uncertainties of the $\pt$ reconstruction from two jets).
The rate for $Z(\to \ell^+\ell^-)+2j$
is smaller than that for $Z(\to \nu\bar\nu) + 2j$ by about a factor of three
due to the relative branching fractions of $Z$ into $ee+\mu\mu$ versus
neutrinos, so the statistics for this measurement will be limited. 
Nevertheless, one can imagine performing a fit or even a subtraction of the
$Z(\to \nu\bar\nu) + 2j$ backgrounds.


\section{The \boldmath $\hin + j$ signal}
\label{sec:hplusjet}

The $\hin + j$ signal comes predominantly from Higgs production via 
gluon fusion, with one radiated jet.  This is the dominant Higgs 
production channel at the Tevatron and the LHC and therefore merits 
attention in our study.     
The production cross section for $h + j$ at a hadron collider was first 
calculated in Ref.~\cite{Ellis:1987xu}.  The cross sections were later
given for the LHC and the Tevatron in Ref.~\cite{Field:2003yy}.  
The total signal cross sections
are approximately 12~pb at the LHC and 0.1~pb at the 
Tevatron for $p_T(j) > 30$~GeV and $|\eta(j)| < 2.5$ \cite{Field:2003yy}.
We calculated the irreducible background for this process from 
$Z (\to \nu {\bar \nu}) + j$ for both LHC and Tevatron using 
Madgraph \cite{Madgraph}.  The cross section for this background is 
$\sim 1.5\times 10^5$~pb at the LHC and $\sim 300$~pb at the Tevatron.  
We see that the number of background events is
larger than that of the signal by a factor of $10^3-10^4$.  The only
handles potentially available to distinguish signal from background are 
the $p_T$ and rapidity of the jet.  However, these distributions are 
similar for the signal and the background, so that significant reduction
of the background is not possible while preserving most of the signal.  
In fact, a mono-jet plus $\pt$ due to  mis-measurement of the jets can
be very substantial, and even impossible to overcome.
Therefore, despite the large signal production rate, $\hin+j$ is not 
a good channel for the discovery of an invisible Higgs.

\section{Discussion and Conclusions}
\label{sec:conclusions}

We studied the signals and backgrounds for an invisibly-decaying Higgs
boson, $\hin$, at present and future hadron colliders.  Such an $\hin$
is motivated by the extremely narrow widths of the SM decay channels
of a Higgs boson below the $W$ pair threshold and the possible existence
of light invisible particles (e.g., dark matter particles or quasistable 
singlets) beyond the
SM to which the Higgs can decay.  For example, in simple models where the 
dark matter particle is a real scalar, the mass of 
this scalar can be as low as $\sim 5$~GeV with ${\cal O}(1)$ couplings to 
the Higgs and $m_h < 150$~GeV \cite{Burgess:2000yq,Davoudiasl:2004be}.  
In the MSSM, an invisible Higgs can arise if $h$ decays predominantly 
to a pair of lightest neutralinos\footnote{While a light enough neutralino 
is disfavored in minimal supergravity due to the chargino mass bound 
from LEP II and the assumption of gaugino mass unification, an 
invisibly-decaying $h$ is a definite possibility in a more general 
MSSM and in extensions of the MSSM containing additional electroweak-singlet 
chiral multiplets~\cite{HLM}.}
or via operators involving goldstinos permitted in non-linearly
realized supersymmetry \cite{Antoniadis}.

In this paper, we have assumed SM production rates for $\hin$ and a 100\%
invisible branching fraction.  Our results can easily be rescaled for 
non-SM Higgs production rates and partly-visible decay branching fractions.
The signal rate is simply
scaled by the production rate and invisible branching fraction:
\begin{equation}
  S = S_0 \frac{\sigma}{\sigma_{SM}} \frac{{\rm BR}_{inv}}{1},
\end{equation}
where $S_0$ is the signal rate from our studies, $\sigma/\sigma_{SM}$ 
is the ratio of the nonstandard production cross section to that of the SM
Higgs, and BR$_{inv}$ is the invisible branching fraction.
Assuming that the SM is the only source of background, the luminosity 
required for a given signal significance then scales like
\begin{equation}
  \mathcal{L} = \mathcal{L}_0 \left[ \frac{\sigma}{\sigma_{SM}}
    \frac{{\rm BR}_{inv}}{1} \right]^{-2},
\end{equation}
where $\mathcal{L}_0$ is the luminosity required for a given significance
found in our studies.

The prospects for the detection of $\hin$ at the LHC 
\cite{Frederiksen:1994me,Godbole:2003it} 
and the Tevatron \cite{Martin:1999qf} in the 
$Z+\hin$ production channel have been studied before.  We revisited 
$Z+\hin$ production at the LHC, including new backgrounds 
not considered in Ref.~\cite{Frederiksen:1994me} and modifying 
the kinematic cuts.  Our results are in good agreement with those
of Ref.~\cite{Godbole:2003it}.
The WBF production of $\hin$ at the LHC has been 
studied in Ref.~\cite{Eboli:2000ze}.  We examined this channel at the 
Tevatron and established the kinematic cuts that are required to detect 
the WBF signal events.  We found that this signal is crucial in making 
the observation of $\hin$ a possibility at the Tevatron.  We also considered 
$\hin + j$ production via gluon fusion at both the Tevatron and LHC.

The $Z+\hin$ channel at the LHC can provide a $\hin$ discovery
with only 10 fb$^{-1}$ for $m_h = 120$ GeV.
With 30 fb$^{-1}$, discovery can be pushed out to $m_h = 160$ GeV.  
This channel can be used at low $m_h$ to supplement
the previously-studied WBF channel, which has higher significance;
it provides a second discovery channel with very different 
experimental systematics to confirm a discovery of $\hin$ in the 
WBF channel.  
The event rates in the $Z+\hin$ and
WBF channels could also be used to extract the Higgs boson mass from the
production cross sections.
Because the $Z+\hin$ cross section falls faster with increasing
$m_h$, it provides more sensitivity than WBF to $m_h$ for 
$m_h \lesssim 160$ GeV once systematic uncertainties are included.
Taking the ratio of rates in $Z+\hin$ and WBF removes dependence on the
production cross section and invisible decay branching fraction, allowing
a more model-independent determination of the Higgs mass, with an 
uncertainty of 35--50 GeV (15--20 GeV) with 10 (100) fb$^{-1}$ of integrated
luminosity.
The $\pt$ distribution is also sensitive
to $m_h$: larger $m_h$ results in a larger average $\pt$ in $Z+\hin$
events.  At the LHC, the production cross section and $\pt$ distribution 
may be the only experimental handles on the mass of a Higgs boson with 
no visible decays.

By itself, $\hin$ production via WBF at the Tevatron provides a 
less than 2$\sigma$ signal even with 10 fb$^{-1}$ of integrated 
luminosity.  However, combining this channel with $Z+\hin$, studied 
previously, and combining the data from the two detectors gives the 
possibility of a 3$\sigma$ observation of a 120 GeV $\hin$ with 7 fb$^{-1}$ 
of delivered luminosity.  This puts 
the observation of an invisibly-decaying Higgs boson within the realm of
possibility at the Tevatron before data from the LHC is available.
Vetoing additional soft jets in the central region could significantly
improve the observability of this channel.

Finally, we observe that $\hin + j$ production via gluon fusion does 
not provide a useful signal at either the Tevatron or LHC.  This is 
because there are not enough handles to separate the signal from the 
overwhelming background.

\acknowledgments

We thank V.~Barger, J.~Conway, I.~Iashvili, F.~Petriello, and A.~Turcot 
for discussions, and R.~Godbole for pointing out Ref.~\cite{Godbole:2003it}.
This work was supported in part by the
U.S.~Department of Energy under grant DE-FG02-95ER40896 and in part by
the Wisconsin Alumni Research Foundation.  The research of H.D. is also 
supported in part by the P.A.M. Dirac Fellowship, awarded by the 
Department of Physics at the University of Wisconsin-Madison.
T.H. is also supported in part  by a Fermilab Frontier Fellowship.
Fermilab is operated by the Universities Research Association 
Inc.~under Contract No.~DE-AC02-76CH03000 with the U.S.~Department of Energy.


\end{document}